\documentclass[fleqn,usenatbib]{mnras}

\usepackage{newtxtext,newtxmath}

\usepackage[T1]{fontenc}
\usepackage{ae,aecompl}

\usepackage{graphicx}	
\usepackage{amsmath}	
\usepackage{amssymb}	

\usepackage{xcolor}

\title[Localization of BBH Mergers with Inclination]{Localization of Binary Black-Hole Mergers with Known Inclination}

\author[K. R. Corley et al.]{%
K. Rainer Corley,$^{1,2}$\thanks{E-mail: rainer.corley@columbia.edu}
Imre Bartos,$^{3}$\thanks{E-mail: imrebartos@ufl.edu}
Leo P. Singer,$^{4,5}$
Andrew R. Williamson,$^{6,7}$
\newauthor
Zolt\'an Haiman,$^{8}$
Bence Kocsis,$^{9}$
Samaya Nissanke,$^{6,7}$
Zsuzsa M\'arka,$^{2}$
and
\newauthor
Szabolcs M\'arka$^{1}$
\\
$^1$Department of Physics, Columbia University in the City of New York, 550 W 120th St., New York, NY 10027, USA\\
$^2$Columbia Astrophysics Laboratory, Columbia University in the City of New York, 550 W 120th St., New York, NY 10027, USA\\
$^3$Department of Physics, University of Florida, PO Box 118440, Gainesville, FL 32611-8440, USA\\
$^4$Astroparticle Physics Laboratory, NASA Goddard Space Flight Center, 8800 Greenbelt Road, Greenbelt, MD 20771, USA \\
$^5$Joint Space-Science Institute, University of Maryland, College Park, MD 20742, USA \\
$^6$GRAPPA, Anton Pannekoek Institute for Astronomy and Institute of High-Energy Physics, University of Amsterdam, Science Park 904, 1098 XH Amsterdam, The Netherlands \\
$^7$Nikhef, Science Park 105, 1098 XG Amsterdam, The Netherlands \\
$^8$Department of Astronomy, Columbia University in the City of New York, 550 W 120th St., New York, NY 10027, USA\\
$^9$E\"otv\"os University, Institute of Physics, P\'azm\'any P. s. 1/A, Budapest, 1117, Hungary
}

\date{Accepted XXX. Received YYY; in original form ZZZ}

\pubyear{2019}

\begin{document}
\label{firstpage}
\pagerange{\pageref{firstpage}--\pageref{lastpage}}
\maketitle

\begin{abstract}
The localization of stellar-mass binary black hole mergers using gravitational waves is critical in understanding the properties of the binaries' host galaxies, observing possible electromagnetic emission from the mergers, or using them as a cosmological distance ladder. The precision of this localization can be substantially increased with prior astrophysical information about the binary system. In particular, constraining the inclination of the binary can reduce the distance uncertainty of the source. Here we present the first realistic set of localizations for binary black hole mergers, including different prior constraints on the binaries' inclinations. We find that prior information on the inclination can reduce the localization volume by a factor of 3. We discuss two astrophysical scenarios of interest: (i) follow-up searches for beamed electromagnetic/neutrino counterparts and (ii) mergers in the accretion disks of active galactic nuclei.
\end{abstract}




\section{\label{sec:introduction}Introduction}

So far, Advanced LIGO \citep{2015CQGra..32g4001L} and Advanced Virgo \citep{2015CQGra..32b4001A} have discovered ten stellar-mass binary black hole (BBH) mergers \citep{2018arXiv181112907T}, a number that is expected to grow by two orders of magnitude in the next few years \citep{2018LRR....21....3A}. Such a high detection rate will soon enable constraining binary formation channels, probing the environments of the mergers, and studying the expansion of the universe, among other possibilities.

Identifying the host galaxies of BBH mergers is particularly useful in these and other endeavors. The properties of host galaxies provide additional information on these events. For example, the association of some BBH mergers with rare host galaxy types---such as active galactic nuclei (AGNs)---can reveal the binaries' formation channel \citep{2017NatCo...8..831B}. Host galaxies are also essential in utilizing BBH mergers as standard cosmological candles \citep{Schutz}.

Beyond the gravitational-wave (GW) signal and host galaxies, further information could be gained of BBH systems if their merger also produces detectable electromagnetic/neutrino counterpart. Such emission can be produced for binaries residing in a dense gaseous environment from which they can accrete at super-Eddington rates. This may be the case in a few astrophysically plausible scenarios, including mergers in the accretion disks of active galactic nuclei \citep{2017ApJ...835..165B,2017MNRAS.464..946S,2017NatCo...8..831B}, gas or debris remaining around the black holes from their prior evolution \citep{2016ApJ...821L..18P,2016ApJ...823L..29K,2016PhRvD..93l3011M,2017ApJ...839L...7D,2016ApJ...822L...9M,2017MNRAS.465.4406K}, and BBH formation inside a collapsing star \citep{2016ApJ...819L..21L,2017MNRAS.470L..92D,2017PhRvL.119q1103F}.

The identification of host galaxies is limited by the localization accuracy of GWs. This accuracy primarily depends on the signal-to-noise ratio of the GW signal, the number of detectors, the masses of the black holes, the sky direction, and any prior information available of the event. 

Prior studies of GW localization have overwhelmingly focused on binary-neutron-star and neutron-star-black-hole mergers (\citealt{2018LRR....21....3A} and references therein), which are known to produce electromagnetic/neutrino emission \citep{2012ApJ...746...48M,2013CQGra..30l3001B,2017ApJ...848L..12A,2017ApJ...850L..35A}. Further, using prior information to improve localization has been considered only recently following the discovery of binary neutron star merger GW170817 \citep{2017PhRvL.119p1101A}. The inclination of GW170817, inferred from the short gamma-ray burst and its afterglow produced by the merger \citep{2019PhRvX...9a1001A,2018Natur.561..355M,2017ApJ...848L..13A}, was greater than what was anticipated for joint gravitational-wave and gamma-ray burst observations. The motivation behind these studies is the degeneracy between orbital inclination and luminosity distance in GW parameter estimation \citep{2010ApJ...725..496N,2015PhRvD..91d2003V,2018arXiv180910727U}. Constraints on the inclination can therefore improve distance measurement, which is the relevant quantity for measuring Hubble's constant with GWs \citep{1986Natur.323..310S}. Improved estimates on the binary inclination can be achieved by combining information from gravitational waves with other electromagnetic observations, in particular a gamma-ray burst afterglow \citep{2018ApJ...869...55W,2018Natur.561..355M}. A previous study \citep{2016arXiv161201471C} also looked the localization volumes of BBH mergers. However, this work only considered representative binary masses and no prior information on inclination.

Knowing the inclination of BBH mergers could help their localization similarly to the case of neutron star mergers. This has, however, not been previously studied as constraining the inclination of BBHs is only possible if more information can be collected about the merger than what is available in GWs. There are two possibilities: electromagnetic emission from the binary through gas accretion from a dense environment, and information from the binary's host galaxy.

In this paper we use inclination priors for stellar-mass BBH mergers in AGN environments to infer the expected probability density of GW localization for realistic mass distributions, for the network of Advanced LIGO and Virgo detectors at design sensitivity, and for a 5-detector network including LIGO-India \citep{LIGOIndia} and KAGRA \citep{PhysRevD.88.043007}. We derive localizations both with and without prior information on localization. We then discuss these results in the context of two relevant search scenarios: (i) the search for a beamed electromagnetic/neutrino counterpart and (ii) BBH mergers in AGN disks. 

\section{Utility of inclination priors}

Having prior information on the inclination of BBH mergers can significantly reduce the gravitational-wave localization volume of the mergers. This reduction has multiple advantages. Below we discuss two such advantages: in follow-up observations searching for the electromagnetic/neutrino counterparts of the merger, and the identification of a sub-class of mergers occurring in AGNs.

\subsection{Electromagnetic/neutrino follow-up}

The search for electromagnetic/neutrino emission from a source can benefit from the use of host galaxy candidates by reducing the part of the sky that needs to be scanned and by reducing the number of false positives \citep{2013ApJ...767..124N,2015ApJ...801L...1B,2017Sci...358.1556C,2016MNRAS.460L..40E,2016ApJ...829L..15S}. A reduction of the search volume therefore proportionally reduces the number of host galaxy candidates that need to be searched for.

We envision two particular scenarios in which prior information on the inclination of the source can be available. First, high-energy gamma-ray and neutrino emission from the source is likely highly beamed. Therefore, if a high-energy counterpart of the GW signal is observed, one can carry out follow-up observations of this joint source by making use of the fact that the orbital axis of the binary is likely pointing close to Earth. Taking cosmological GRBs as an example, high-energy emission will be visible within $\lesssim10^\circ$ \citep{2014ARA&A..52...43B,2016ApJ...818...18G}. While in the case of GW170817/GRB\,170817A, high-energy emission has been observed at about 20$^{\circ}$ \citep{2018Natur.561..355M,2018arXiv180800469G}, this will probably only be the case for the most nearby events \citep{2017ApJ...848L..13A,2017ApJ...848L..14G,2018arXiv180806238G}, which will be known in advance from the GW signal. A possible caveat is that jets driven by accretion disks around orbiting black holes may be precessing with the motion of the black hole, producing high-energy emission in a changing direction that is offset from the orbital axis \citep{2010Sci...329..927P,2018PhRvD..97d4036K,2017A&A...602A..29B}. This can introduce an additional offset over the jet opening angle, however this will not be significantly increase the $\lesssim10^\circ$ offset.

As a second possibility, we envision a merger whose orbital axis is pointing away from Earth. While high-energy emission is not detected from such an event, if the merger drives a relativistic outflow it will produce afterglow emission that spreads to greater angles. The profile of this off-axis afterglow emission is dependent, among other parameters, on time \citep{2011ApJ...733L..37V}. Therefore, for a given time, one can estimate the inclinations for which the afterglow emission should be observable at that time, making it possible to optimize the follow-up observation schedule of host galaxy candidates. For example, there can be some galaxies within the gravitational wave localization volume that are too far to be host galaxies unless the binary inclination is close to zero. These galaxies should be looked at first as afterglow has no time delay. Some other galaxies that are closer may only be host galaxies if the binary inclination is large. For large inclinations, afterglow should be delayed, and therefore these galaxies are only worth following up with a time delay after the merger.

\subsection{Inclination prior from AGNs}

Active galactic nuclei represent a special promising case of BBH mergers in which the inspiral of the BHs is accelerated and guided by the AGN disk \citep{2012MNRAS.425..460M,2017ApJ...835..165B,2017MNRAS.464..946S,2019ApJ...876..122Y,2019arXiv190609281Y}. As the BHs migrate into the AGN disk's plane, the binary orbital axis becomes aligned with the AGN disk's orbital axis. Therefore, information on the orientation of the AGN disk will inform the expected inclination of the BBH merger.

The orientation of the AGN disk may be measured with several distinct methods. This includes fitting the observed iron K$\alpha$ line profile with accretion disk models \citep{1997ApJ...477..602N}; using the broadening of the H$\beta$ emission line with known SMBH mass from, e.g., reverberation mapping \citep{2001ApJ...561L..59W} (for the inclination of the Broad-Line Region); polarization measurements \citep{2003A&A...404..505S}; and others (see \citealt{2014MNRAS.441..551M} and references therein). Some of these methods produced inclination estimates with uncertainty $\lesssim 5^\circ$ for some of the studied AGNs (e.g., \citealt{2014MNRAS.441..551M}).

The above information can utilized as follows. We first localize a detected BBH merger using no prior information. Then, within its localization volume, we identify AGNs, either using a galaxy catalog or by follow-up observations \citep{2015ApJ...801L...1B}. For each of these AGNs, we determine the AGN disk's inclination with a suitable technique, and localize the GW signal with this reconstructed inclination as prior. An AGN will only remain a candidate if it remains within the new localization volume derived using the AGN's inclination. 

\section{\label{sec:methods}Methods}

Our methods include generating a sample of realistic BBH merger events and localizing them with various priors on inclination.  Below, we discuss BAYESTAR, the localization algorithm used in this study, and our method of computing BBH localization volumes.

\subsection{BAYESTAR}

We performed our localizations of simulated BBHs using BAYESTAR \citep{2016PhRvD..93b4013S}. BAYESTAR performs a Bayesian analysis using the maximum-likelihood parameter estimates from matched filter searches to compute GW localization.

Previously, BAYESTAR computed posteriors assuming a uniform distribution in $\cos(\iota)$ within [$-1$, $1$], where $\iota$ is the inclination angle (see Eq. (38) in \citealt{2016PhRvD..93b4013S}). We modified the analysis method to include prior information on the binary inclination. In the following we use prior distributions that are uniform in $\cos(\iota)$ within [$\cos(\iota_\mathrm{high})$, $\cos(\iota_\mathrm{low})$]. 

\subsection{Computation of BBH localization volumes}

We adopted a realistic BBH mass model in which each BH mass follows $m\propto m^{-2.35}$, with a minimum mass of $5\,M_\odot$ and a maximum mass of $50\,M_\odot$ \citep{2016PhRvX...6d1015A}. We did not consider any correlation between the two masses. We drew the locations of BBH mergers assuming a uniform distribution in comoving volume, and we assigned a random orientation to the binary.  We used reduced-order-model (ROM) SEOBNRv4 waveforms \citep{2017PhRvD..95d4028B}. We assumed black hole spins aligned with the orbital axis, and we did not include precession. We adopted cosmological parameters from the nine-year WMAP observations \citep{2013ApJS..208...19H}. 

We consider two network sensitivities: (i) the Advanced LIGO and Virgo detectors (HLV), all at design sensitivity \citep{2018LRR....21....3A}, and (ii) LIGO-Virgo-India-KAGRA network (HLVIK), with HLV at A+ sensitivity \citep{T1800042} and IK at design sensitivity.  We assigned each detector an independent duty cycle of 70\%.

We generate samples of BBH parameters as described above and determine whether each injection is detectable. We compute the time of arrival, phase of arrival, and signal-to-noise ratio (SNR) in each detector, and we add Gaussian measurement error to the SNR. We calculate network SNR as the quadrature sum of the individual SNRs, and to consider an event discovered, we use a threshold network SNR of 12 and require at least two detectors to contribute an SNR of 4. For discovered signals, we use BAYESTAR to derive 3D localizations, also using our inclination priors when applicable.

\section{Results}
We discuss two results of the localization studies below: first, the reduction in localization volumes obtained using inclination priors, and second, the improved study of AGNs as a BBH formation channel.
\begin{figure}
    \centering
    \includegraphics[width=\linewidth]{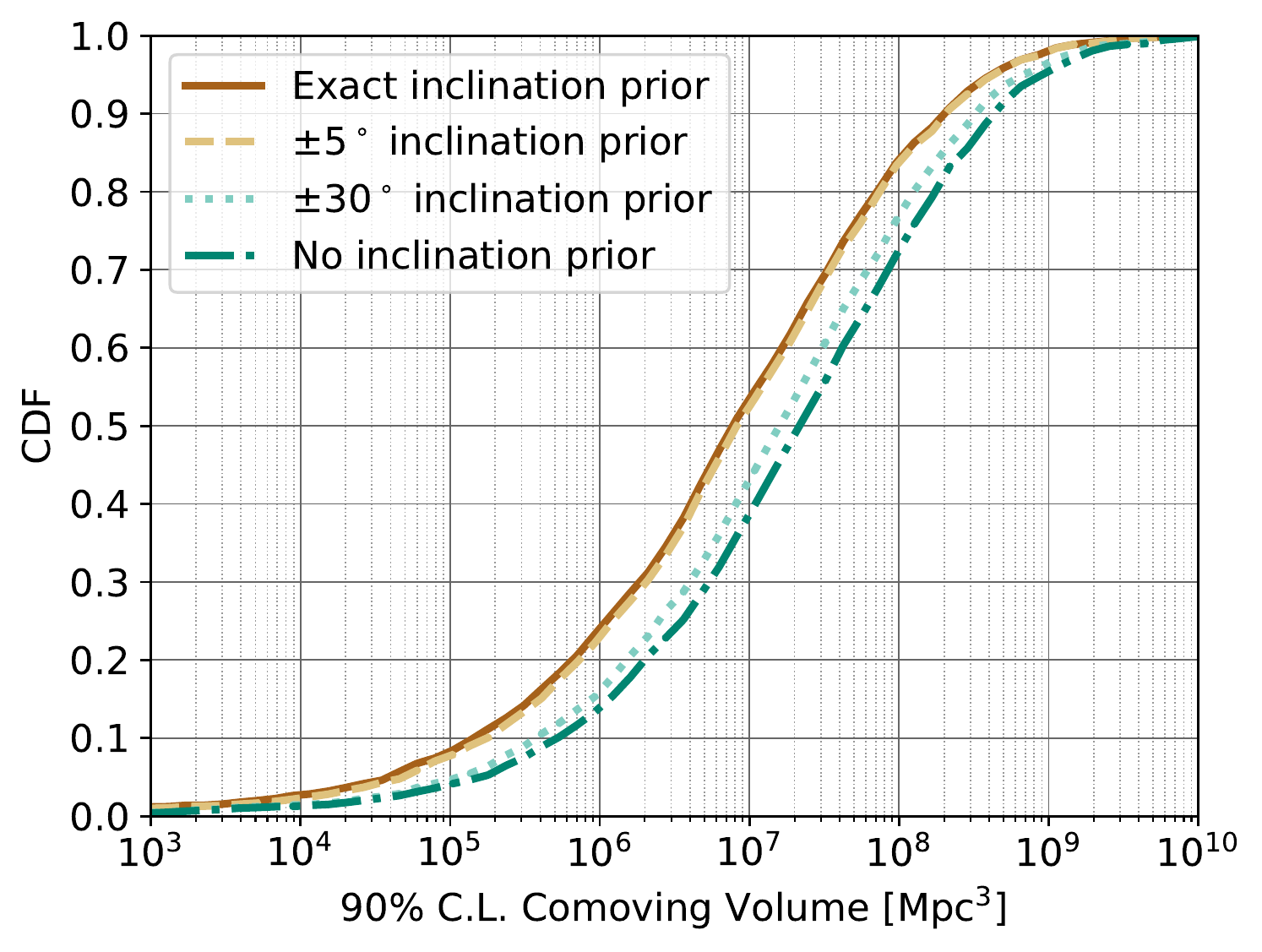}
    \caption{\label{fig:priors_volumes} Cumulative density function (CDF) of the 90\% confidence-level localizations in comoving volume for BBHs detected by the HLV network at design sensitivity, assuming various inclination priors. We find similar distributions for the HLVIK network, due to the simultaneous effects of greater source distances and improved sky localization.}
\end{figure}

\subsection{Localization distribution with inclination priors}

We used the simulations described above to obtain the cumulative probability density of BBH merger localization volumes. Fig. \ref{fig:priors_volumes} shows how localization depends on prior information on the binary inclination. We see that knowing the inclination to $\sim 5^\circ$ precision or better (e.g., \citealt{2014MNRAS.441..551M}) reduces the localization volume by a factor of about 3, while knowing the inclination to a precision of $\sim30^\circ$ reduces the localization volume by about 35\%. 

While not shown separately, we find that this improvement comes from better distance estimation. The improvement of distance estimation is practically the same as the improvement in 3D localization. The 2D sky localization is practically unaffected.

\subsection{Population constraints on AGN-assisted BBH mergers}

We can convert the expected improvement using inclination priors into improved population constraints on the subset of BBH mergers occurring in AGN disks. For this we adopted the method of \citealt{2017NatCo...8..831B}, who used the correlation of AGN locations and GW localization volumes to derive the number of GW detections needed to statistically indicate the presence of an AGN-assisted BBH sub-population. We carried out their calculation with our localization volume distribution assuming that the inclination of the merger is known to within 5$^\circ$.

\begin{figure}
    \centering
    \includegraphics[width=\linewidth]{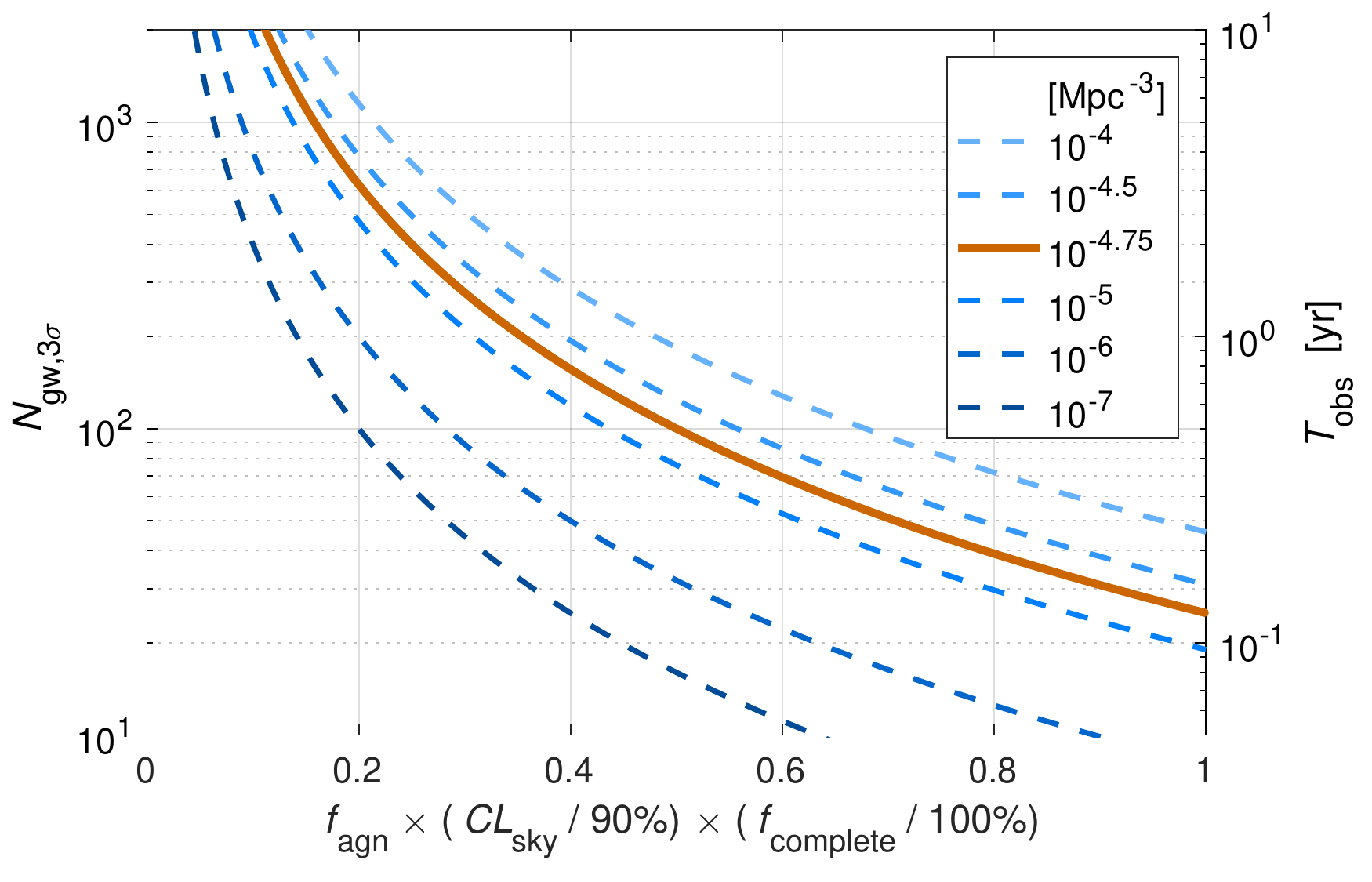}
    \caption{Number of BBH detections required to obtain 3$\sigma$ evidence for the fraction of BBH mergers originating in AGNs with known inclination priors, at various galaxy number densities.\label{fig:BBHAGNdetectiontime}}
\end{figure}

The number of mergers needed for identifying a sub-population as a function of the sub-population's fractional contribution to the total event rate is shown in Fig. \ref{fig:BBHAGNdetectiontime}. Comparing these results to those of \citealt{2017NatCo...8..831B}, we find that the number of events needed is reduced by about a factor of 3 (i.e., proportionally to the improvement in localization). 

\subsection{The role of BH spins}

For AGN-assisted mergers, large BH spins that are misaligned from the orbital axis can mitigate the precision of this technique. In such cases the binary undergoes orbital precession, which can reduce the alignment of the orbital plane from the AGN disk plane. 

In addition, for some precessing binaries, precession can help reconstruct the inclination of the orbit, decreasing the advantage of having additional information on the inclination through other means. However, it is difficult to substantially constrain pre-merger inclination for precessing binaries using Earth-based gravitational-wave instruments \citep{2012PhRvD..86f4020B,2017PhRvD..95f4053V}, unless the BH masses are low \citep{2014PhRvL.112y1101V}. This is due to the fact that lower-mass BHs undertake more orbits and therefore more precession in the LIGO band.

In our studies, we neglected these effects as the spin distribution and alignment of merging black holes is highly uncertain, with no evidence so far for a sufficiently high spin to necessitate accounting for precession \citep{2018arXiv181112907T}. Nevertheless, we note that if a large fraction of the merging black holes had large misaligned spins, it would need to be accounted for in inclination studies.

\section{Conclusions}

We computed the expected probability distribution of GW localization volumes of stellar-mass BBH mergers using a new localization algorithm BAYESTAR$^{\rm i}$, using prior information on the inclination of the binary. We examined two particular cases in which inclination priors can be expected: electromagnetic emission and AGN-assisted mergers. Our takeaways are as follows:
\begin{itemize}
\item Knowing the inclination of the merger a priori to at least 5$^\circ$ reduces the localization volume by about a factor of 3. This factor comes primarily from the reduced distance uncertainty. Knowing the inclination to 30$^\circ$ results in a 35\% reduction.
\item Inclination priors can be used in electromagnetic follow-up searches when beamed high-energy emission or its afterglow is searched for.
\item The inclination of BBH mergers in AGNs can be constrained by assuming that the merger plane is aligned with the AGN disk plane. Using these aligned inclinations can reduce the number of BBH detections needed to identify an AGN-assisted sub-population by a factor of 3.
\end{itemize}

The results above represent a first step in the development of methods to incorporate astrophysical prior information in the localization of BBH mergers.

\section*{Acknowledgments}
The authors thank Richard O'Shaughnessy and Raymond Frey for their valuable feedback. The authors are thankful for the generous support of the University of Florida and Columbia University in the City of New York. The Columbia Experimental Gravity group is grateful for the generous support of the National Science Foundation under grant PHY-1708028. ZH acknowledges financial support from NSF via grant 1715661 and from NASA via grants NNX17AL82G and 80NSSC19K0149. This project has received funding from the European Research Council (ERC) under the European Union's Horizon 2020 research and innovation programme ERC-2014-STG under Grant Agreement No. 638435 (GalNUC) and from the Hungarian National Research, Development, and Innovation Office Grant No. NKFIH KH-125675 (to BK).

This work made use of the following software: Astropy \citep{2013A&A...558A..33A,2018AJ....156..123A}, HEALPix \citep{2005ApJ...622..759G}.




\bibliographystyle{mnras}
\bibliography{BBH-inclination} 








\bsp	
\label{lastpage}
\end{document}